\documentclass[sn-mathphys-num]{sn-jnl}% Math and Physical Sciences Numbered Reference Style 
\usepackage{graphicx}%
\usepackage{multirow}%
\usepackage{amsmath,amssymb,amsfonts}%
\usepackage{amsthm}%
\usepackage{mathrsfs}%
\usepackage[title]{appendix}%
\usepackage{textcomp}%
\usepackage{manyfoot}%
\usepackage{booktabs}%
\usepackage{algorithm}%
\usepackage{algorithmicx}%
\usepackage{algpseudocode}%
\usepackage{listings}%
\usepackage{tabularx}
\usepackage[table,xcdraw]{xcolor}
\usepackage{lmodern}
\usepackage{anyfontsize}
\usepackage{longtable}
\usepackage{tabu}
\usepackage{tikz}
\usetikzlibrary{fit, calc, decorations.markings}

\begin{document}
\begin{sloppypar}

\title[Article Title]{Artificial Intelligence for Central Dogma-Centric Multi-Omics: Challenges and Breakthroughs}


%%作者信息
\author[1,2]{\fnm{Lei} \sur{Xin}}
\equalcont{These authors contributed equally to this work.}

\author[3]{\fnm{Caiyun} \sur{Huang}}
\equalcont{These authors contributed equally to this work.}

\author[4,5,6]{\fnm{Hao} \sur{Li}}
\equalcont{These authors contributed equally to this work.}

\author[7]{\fnm{Shihong} \sur{Huang}}
\equalcont{These authors contributed equally to this work.}

\author[8]{\fnm{Yuling} \sur{Feng}}

\author[9]{\fnm{Zhenglun} \sur{Kong}}

\author[10]{\fnm{Zicheng} \sur{Liu}}

\author[10]{\fnm{Siyuan} \sur{Li}}

\author[10]{\fnm{Chang} \sur{Yu}}

\author[11]{\fnm{Fei} \sur{Shen}}

\author*[1]{\fnm{Hao} \sur{Tang}}\email{haotang@pku.edu.cn}

\affil*[1]{\orgdiv{School of Computer Science}, \orgname{Peking University}, \orgaddress{\city{Beijing}, \postcode{100871}, \country{China}}}

\affil[2]{\orgname{Wuhan University}, \orgaddress{\city{Wuhan}, \postcode{430072}, \state{Hubei}, \country{China}}}

\affil[3]{\orgdiv{Interdisciplinary Institute of Life Medicine}, \orgname{Hunan University}, \orgaddress{\city{Changsha}, \postcode{410082}, \state{Hunan}, \country{China}}}

\affil[4]{\orgdiv{Laboratory of Data Intelligence and Cross Innovation}, \orgname{Nanjing University}, \orgaddress{\city{Nanjing}, \postcode{210023}, \state{Jiangsu}, \country{China}}}

\affil[5]{\orgdiv{School of Information Management}, \orgname{Nanjing University}, \orgaddress{\city{Nanjing}, \postcode{210023}, \state{Jiangsu}, \country{China}}}

\affil[6]{\orgdiv{College of Information Management}, \orgname{Nanjing Agricultural University}, \orgaddress{\city{Nanjing}, \postcode{210095}, \state{Jiangsu}, \country{China}}}

\affil[7]{\orgdiv{Faculty of Science}, \orgname{Nanjing Agricultural University}, \orgaddress{\city{Nanjing}, \postcode{210095}, \state{Jiangsu}, \country{China}}}

\affil[8]{\orgdiv{Department of Pharmacology}, \orgname{University of Michigan Medical School}, \orgaddress{\city{Ann Arbor}, \postcode{48109}, \state{Michigan}, \country{USA}}}

\affil[9]{\orgdiv{Department of Biomedical Informatics}, \orgname{Harvard Medical School}, \orgaddress{\city{Boston}, \postcode{02115}, \state{Massachusetts}, \country{USA}}}

\affil[10]{\orgdiv{AI Lab, Research Center for Industries of the Future }, \orgname{Westlake University}, \orgaddress{\city{Hangzhou}, \postcode{310058 }, \state{Zhejiang}, \country{China}}}

\affil[11]{\orgdiv{Beijing Key Laboratory of Agricultural Genetic Resources and Biotechnology, Institute of Biotechnology}, \orgname{Beijing Academy of Agriculture and Forestry Sciences}, \orgaddress{\city{Beijing}, \country{China}}}


\abstract{
With the rapid development of high-throughput sequencing platforms, an increasing number of omics technologies, such as genomics, metabolomics, and transcriptomics, are being applied to disease genetics research. However, biological data often exhibit high dimensionality and significant noise, making it challenging to effectively distinguish disease subtypes using a single-omics approach. To address these challenges and better capture the interactions among DNA, RNA, and proteins described by the central dogma, numerous studies have leveraged artificial intelligence to develop multi-omics models for disease research. These AI-driven models have improved the accuracy of disease prediction and facilitated the identification of genetic loci associated with diseases, thus advancing precision medicine.
This paper reviews the mathematical definitions of multi-omics, strategies for integrating multi-omics data, applications of artificial intelligence and deep learning in multi-omics, the establishment of foundational models, and breakthroughs in multi-omics technologies, drawing insights from over 130 related articles. It aims to provide practical guidance for computational biologists to better understand and effectively utilize AI-based multi-omics machine learning algorithms in the context of central dogma.}

%%================================%%
%% Sample for structured abstract %%
%%================================%%

\maketitle

\section{Core Issues of Multi-Omics and Bottleneck Problems in Biological Experiments}\label{sec1}

\subsection{Concept and Advantages of Multi-Omics}\label{subsec11}
Multi-omics refers to the integrated analytical approach combining single-omics fields such as genomics, transcriptomics, proteomics, metabolomics, and microbiomics. While single-omics approaches offer specific advantages, they have inherent limitations when it comes to comprehensive analysis. For example, genomics and transcriptomics struggle to effectively link genotype to phenotype \cite{1,2}. Genome-Wide Association Studies (GWAS), widely used in genomics, can identify loci associated with human disease traits but cannot pinpoint target genes, cell types, or specific biological functions. This limitation reduces their applicability in clinical and precision medicine \cite{3}. Similarly, metabolomics faces challenges; in typical untargeted metabolomics studies of human samples, only about 10\% of the data can be structurally annotated \cite{4,5,6,7}.
The strength of multi-omics lies in its ability to complement the shortcomings of individual omics technologies. By integrating diverse genomic patterns \cite{8}, multi-omics enables in-depth cellular analyses, uncovering genomic and transcriptomic changes during growth, differentiation, or pathogenesis \cite{9}. For instance, studies combining proteomics with genomic and transcriptomic data have revealed the critical role of driver genes in cancer \cite{10}. Complex diseases often exhibit temporal heterogeneity within individuals, and multi-omics provides a powerful approach for comprehensive phenotypic analysis of transitions from health to disease. This facilitates precision medicine and the development of targeted treatments.

\subsection{Applications of Multi-Omics in Disease Research}\label{subsec12}
Omics technologies encompass various modalities, each tied to the central dogma of molecular biology: DNA → RNA → protein. The base composition of DNA, regulated by adenine (A), thymine (T), guanine (G), and cytosine (C), plays a key role in transcriptional regulation. For example, changes in GC content are associated with transcriptional activity. In vertebrates, GC content decreases progressively along the gene from the 5' untranslated region (UTR) to the 3' UTR, whereas in invertebrates, coding regions exhibit the highest GC content \cite{11}.
While the entire genome can theoretically be transcribed, 98\% of the human genome is transcribed into thousands of non-coding RNAs (ncRNAs) \cite{12,13}. Unlike protein-coding RNAs, ncRNAs do not encode proteins \cite{14}. However, many ncRNAs play vital roles in regulatory processes. For instance, certain ncRNAs are involved in translational regulation \cite{15}. In genome-wide association studies (GWAS) identifying susceptibility variants for common diseases, the non-coding RNA 7SK has been shown to regulate transcriptional stability and termination in vertebrates \cite{16,17}.
Long non-coding RNAs (lncRNAs) can recruit chromatin regulatory factors to specific genomic regions, thereby influencing gene expression \cite{18}. Short ncRNAs, such as microRNAs, also play a critical role in transcriptional regulation \cite{19,20}. MicroRNAs mediate post-transcriptional gene silencing by guiding Argonaute (AGO) proteins to the 3' UTR of target mRNAs, effectively inhibiting gene expression \cite{21,22}.

\subsection{Function and Importance of Non-Coding Area}\label{subsec13}
Single-omics approaches cannot explain how identical DNA sequences result in different expression patterns across various cell types. However, multi-omics data can provide comprehensive analyses to address these challenges. Zhang et al. used advanced multi-omics techniques to investigate genomes, transcriptomes, and functional proteomes, uncovering that the extraordinary radiation resistance of tardigrades is attributed to 2,801 genes that respond rapidly to radiation \cite{23}.
Multi-omics transcriptional regulation is a highly complex process influenced by numerous regulatory factors, including transcription factors (TFs) and modifications such as methylation, phosphorylation, and acetylation \cite{24,25}. In practical medical settings, many diseases—such as cancer, kidney disease, autoimmune disorders, and cardiovascular diseases—can arise from mutations in transcription factors, regulatory sequences, or non-coding RNAs \cite{26,27,28}. Transcription factors regulate gene transcription by binding to specific DNA sequences, regulatory sequences influence the spatiotemporal specificity of gene expression, while non-coding RNAs affect cellular functions by regulating gene expression and participating in processes such as RNA splicing and translation. For instance, dysregulation of DNA methylation has been implicated in cancer development \cite{29}, while research by Wang et al. \cite{30} found that 14\% of gastric cancer patient samples carry genetic alterations in the TEAD gene.

\subsection{Challenges in Multi-Omics Data Analysis}\label{subsec14}
The primary goal of multi-omics analysis is to extract as much valuable information as possible from its complex biological data. With the rapid advancement of sequencing technologies, multi-omics sequencing has become increasingly sophisticated. Platforms such as the Illumina HiSeq X analyzer and the Visium platform are capable of performing multi-omics sequencing effectively \cite{31}.
However, the storage formats and structures of multi-omics biological data vary significantly, resulting in substantial isolation between different omics data types. Extracting meaningful insights across these diverse data types is challenging. For instance, uncovering hidden relationships between genomic transcriptional expression and phenotypic data often requires complex nonlinear modeling, which leads to high-dimensional data. Furthermore, such data may be noisy or imperfectly matched, adding to the difficulty of analysis \cite{32}.
Another persistent challenge in multi-omics biology is the difficulty of spatially aligning in vivo imaging data with ex vivo data \cite{33}. Additionally, multi-omics data from different institutions often exhibit isolation and heterogeneity, which is a key factor contributing to the failure of AI systems in clinical experiments \cite{34}.

\begin{figure}[tbp]
    \centering
    \includegraphics[width=1\textwidth]{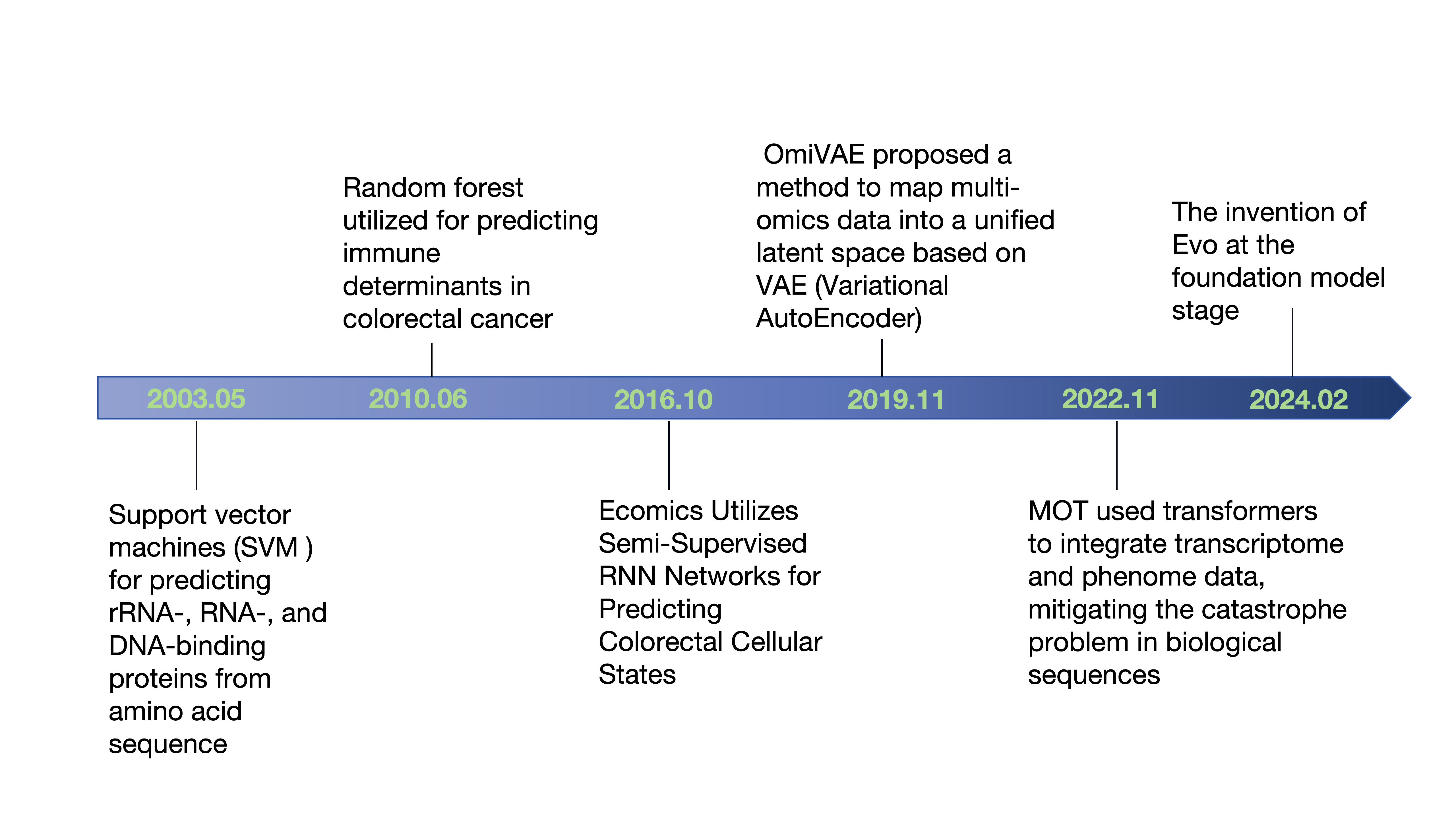}
    \caption{\textbf{Timeline of AI-driven multi-omics methodologies.} We select representative technologies (SVM\cite{fr1}, random forest\cite{fr2}, RNN\cite{fr3}, VAE\cite{fr4}, Transformer\cite{fr5}, LLM\cite{fr6}) and identified their emergence in multi-omics as key milestones for constructing the timeline.}
    \label{fig1}
\end{figure}

\subsection{Future of High-Throughput Technology and Data Sharing}\label{subsec15}
The analysis of multi-omics biological data faces significant challenges, including data sparsity and noise. For instance, biological multi-omics data obtained through the 10x Genomics platform often exhibit these issues \cite{35}. Factors such as sample absence, complex data structures, heterogeneous categories, and dataset shifts contribute to data inaccessibility and sparsity, complicating multi-omics sequencing and analysis \cite{36}.
To address these challenges, there is a pressing need for more effective methods and tools capable of extracting meaningful signals from noisy data and tackling data scarcity. Some studies have proposed a distribution-based algorithm for feature selection and distribution (FSD) combined with four machine learning methods—MLP, RF, XGBoost, and SVM—to reduce dimensionality and filter noise in multi-omics datasets \cite{37}.
The growing maturity of high-throughput technologies and increased data sharing between institutions can partially mitigate the limitations of data sparsity in multi-omics research. However, the high dimensionality, noise, sparsity, and isolation inherent in multi-omics data remain significant obstacles to effective analysis \cite{38}.
With advancements in artificial intelligence and deep learning, these technologies offer promising solutions to the challenges of high dimensionality, noise, and data sparsity in multi-omics. Leveraging AI methods can improve the application of multi-omics in precision medicine, supporting efforts in disease prevention, diagnosis, and personalized treatment.

\section{Multimodel Alignment Strategies\label{sec2}}
There are notable differences in the vocabulary and biological prior knowledge associated with different biological macromolecules. Most existing multi-omics models are trained on single modalities and align their dimensions using linear or nonlinear connectors, employing suitable strategies for data fusion. In this context, we present three mainstream post-fusion strategies for integrating biological macromolecules:

\begin{enumerate}
\item \textbf{Linear Projector} 
is a classic method for multimodal alignment, efficiently aligning the dimensions of outputs from three types of biological macromolecules within the network using a single linear layer.
\begin{equation}
Y = WX + B,
\end{equation}
where \( Y \) is the output, \( W \) is the weight matrix, \( X \) is the input, and \( B \) is the bias term. The linear layer is commonly used for feature mapping and dimensionality reduction, effectively combining input features linearly to prepare for subsequent nonlinear activation.

\item \textbf{Multi-Layer Perception (MLP)} 
is a neural network architecture consisting of multiple fully connected layers, designed to account for the influence of nonlinear factors during multimodal alignment. To effectively capture complex patterns and relationships in the data, MLP models typically include two feedforward layers combined with activation functions, which act as feature extraction units. During processing, the dimensions of the feedforward layers first expand and then contract, enabling the model to focus on capturing specific data features.
\begin{equation}
Y = f(W_2 \cdot f(W_1 \cdot X + B_1) + B_2),
\end{equation}
where \( f \) is the activation function (e.g., ReLU), \( W_1 \) and \( W_2 \) are the weights, and \( B_1 \) and \( B_2 \) are the biases. MLPs are widely used for feature learning and classification tasks, capturing complex feature patterns through their multi-layer structure.

\item \textbf{Cross-Attention Mechanism}, 
originally introduced in the Transformer \cite{66} architecture, is designed to integrate information from different sequences. In the context of multi-omics sequence modeling, it effectively captures interactions among the three types of biomolecules: DNA, RNA, and proteins.
\begin{equation}
\text{Attention}(K, Q, V) = \operatorname{softmax}\left(\frac{QK^T}{\sqrt{d_k}}\right){V}.
\end{equation}
In multimodal learning, the queries \( Q \), keys \( K \), and values \( V \) 
can originate from different modalities, enabling effective alignment and integration of information across them.

\item \textbf{Q-Former} 
is a lightweight Transformer architecture designed to convert variable-length inputs into fixed-length query representations\cite{79}.Initially developed for visual-text cross-modal alignment tasks, it is now applied to integrate multi-omics data for biological macromolecules. In the Q-Former module, learnable query embeddings interact with input features via multi-head attention and cross-attention, facilitating effective information exchange.
\begin{equation}
G_r \left[ I_{in}(i,j) - I_{in}(m,n) \right] \exp \left\{ -\frac{\left[ I_{in}(i,j) - I_{in}(m,n) \right]^2}{2\sigma_r^2} \right\},
\end{equation}
where \( G_r \) denotes a specific feature extraction or weighting method. \( I_{in}(i,j) \) and \( I_{in}(m,n) \) represent the feature values of the input image at different positions. 
This formula computes the difference between the feature values and incorporates a Gaussian function to evaluate the relevance and similarity of the features.

\end{enumerate}

\section{Applications of Deep Learning Methods in Multi-Omics}\label{sec3}

\subsection{Classification Tasks}\label{subsec31}
Traditional machine learning methods, such as support vector machines (SVM), were widely used in early multi-omics classification tasks \cite{39,40}. For example, Ronesh Sharma et al. utilized SVM to predict DNA, RNA, and protein binding sites, achieving an accuracy of nearly 80\% \cite{39}. However, as a linear model, SVM struggles to capture the nonlinear dependencies between different biological features, limiting its effectiveness in integrating complex multi-omics data.

In recent years, deep learning models have excelled in multi-omics classification tasks due to their ability to capture nonlinear relationships, establishing them as a mainstream research direction. Sequence models, such as Long Short-Term Memory networks (LSTM), Bidirectional LSTM, and Gated Recurrent Units (GRU), are classic deep learning architectures widely applied in the multi-omics field. For instance, Faisal Ghaffar et al. achieved an accuracy of 90.419\% using GRU on a custom dataset \cite{41}.
However, sequence models face limitations in handling long-distance dependencies and multimodal fusion challenges. Consequently, researchers have increasingly explored other deep learning models for classification tasks, with convolutional architectures emerging as a prominent solution. For example, the Convolutional Graph Neural Network (ConvGNN) improved classification accuracy in COPD tasks by approximately 6.36\% compared to traditional machine learning methods \cite{42}.
Another notable approach involves Graph Neural Network (GNN)-based models \cite{43,44,45}, such as the Graph Attention Network (GAT). GAT enhances GNN by incorporating an attention mechanism, combining the ability to integrate diverse data types through graph structures with the capacity to capture critical relationships via attention mechanisms. A notable example is LASSO-MOGAT, developed by Fadi Alharbi, which applied GAT to classify multi-omics cancer datasets from The Cancer Genome Atlas (TCGA). By integrating multi-omics data, including mRNA, miRNA, and DNA methylation, the model achieved a classification accuracy of 94.68\%, representing a 4.69\% improvement over traditional Graph Convolutional Network (GCN) models \cite{43}.

When handling longer sequence data, transformer-based architectures offer distinct advantages. These benefits stem from the self-attention mechanism and the powerful feature-learning capabilities of transformer models, which are particularly effective for multimodal data fusion in long sequences. For instance, Liangrui Pan's Supervised Multi-head Attention Mechanism (SMA) model achieved outstanding results on multi-omics datasets from TCGA, including breast cancer, glioblastoma, sarcoma, lung adenocarcinoma, and gastric cancer, with an average classification accuracy of 99.16\% \cite{46}.

Innovations in neural network architectures have also produced remarkable results. For example, Asim Waqas et al. introduced self-normalizing activation functions (SELU) into neural network design, leading to the development of the Self-Normalizing Multi-Omics Model (SeNMo). This model achieved an impressive accuracy of 99.8\% in classifying 33 cancer types from TCGA, showcasing its strong capability in handling high-dimensional data and multimodal data fusion \cite{47}.
While transformer models excel in classification tasks, their quadratic complexity in processing long sequences presents a significant challenge. Thus, achieving a balance between computational overhead and performance is critical for long biological sequence modeling problems.

\begin{tiny}
\begin{longtable}{p{1.5cm}p{1cm}p{2cm}p{1.5cm}p{2.5cm}p{1cm}p{0.5cm}}

\caption{\textbf{Application of modern machine learning techniques in multi-omics classification tasks. We list them according to their categories.}}
\label{tab1} \\
\toprule 
    \rowcolor[HTML]{C6CCD9}
    \textbf{Specific Methods} & \textbf{Model Categories} & \textbf{Specific Classification tasks} & \textbf{Evaluation Indicators} & \textbf{Data Sources} & \textbf{Venue} & \textbf{Year} \\ \midrule
    \rowcolor[HTML]{F2F3F6} 
    DRPBind\cite{39} & SVM & Classification of binding sites & Acc: 76\% & Benchmark dataset from Zhang et al. & bioRxiv & 2023 \\ \hline
    \rowcolor[HTML]{E1E4EB}
    Substrate PHosphosite-based Inference for Network of KinaseS (SPHINKS) \cite{40} & SVM & Classify glioblastoma (GBM) into four functional subtypes based on multi-omics data & AUROC: 87\% & CPTAC-GBM dataset, TCGA-GBM cohort & Nature Cancer & 2023 \\ \hline
    \rowcolor[HTML]{F2F3F6} 
    Ghaffar et al. \cite{41} & Bi-LSTM & Macromolecule classification & Acc: 77.54\% & Macromolecule sequence dataset & Research-Gate & 2020 \\ \hline
    \rowcolor[HTML]{E1E4EB}
    Ghaffar et al. \cite{41} & CNN-LSTM & Macromolecule classification & Acc: 79.64\% & Macromolecule sequence dataset & Research-Gate & 2020 \\ \hline
    \rowcolor[HTML]{F2F3F6} 
    Ghaffar et al. \cite{41} & GRU & Macromolecule classification & Acc: 90.42\% & Macromolecule sequence dataset & Research-Gate & 2020 \\ \hline
    \rowcolor[HTML]{E1E4EB}
    Ghaffar et al. \cite{41} & CNN & Macromolecule classification & Acc: 98.8\% & Macromolecule sequence dataset & Research-Gate & 2020 \\ \hline
    \rowcolor[HTML]{F2F3F6} 
    Convolutional Graph Neural Network (ConvGNN) \cite{42} & CNN/GNN & Classification of Chronic Obstructive Pulmonary Disease (COPD) & Acc: 74.86\% & COPDGene study & PLOS ONE & 2023 \\ \hline
    \rowcolor[HTML]{E1E4EB}
    LASSO-Multi-Omics Gated Attention (LASSO-MOGAT) \cite{43} & GNN & Classification of 31 cancer types & Acc: 94.68\% & TCGAbiolinks library & Academia Biology & 2024 \\ \hline
    \rowcolor[HTML]{F2F3F6} 
    Heterogeneous Graph ATtention network for omics integration (HeteroGATomics) \cite{44} & GNN & Binary and multi-categorical tasks for three cancer types & Acc: 96.4\% & TCGA & arXiv & 2024 \\ \hline
    \rowcolor[HTML]{E1E4EB}
    Adaptive Conditional Graph Diffusion Convolution (ACGDC) \cite{45} & GNN & Classification and identification of astrocyte subtypes based on single-cell data & Sim: 58.84\% & The GTEx (Genotype-Tissue Expression) Analysis V8 release & bioRxiv & 2023 \\ \hline
    \rowcolor[HTML]{F2F3F6} 
    Supervised Multi-head Attention Mechanism model (SMA)  \cite{46} & Trans-former & Identification of cancer subtypes & Acc: 99.16\% & TCGA & IEEE & 2023 \\ \hline
    \rowcolor[HTML]{E1E4EB}
    Self-normalizing Network for Multi-omics (SeNMo) \cite{47} & Self-Normaliz-ing Neural Network & Prediction of primary cancer type & Acc: 99.8\% & Genomics Data Commons (GDC) & arXiv & 2024 \\ \bottomrule
\end{longtable}
\end{tiny}

\subsection{Regression Tasks}\label{subsec32}
In recent years, deep learning models have been extensively applied to multi-omics tasks, including gene expression prediction, drug response prediction, and modeling cellular states under specific conditions or perturbations \cite{52, fr3}. These frameworks integrate data from diverse sources, such as DNA methylation, RNA sequencing, and proteomics, effectively uncovering complex molecular mechanisms.
For example, Wang et al. employed deep neural networks with attention layers to integrate various omics features for drug response prediction, emphasizing the critical role of molecular-level information, such as gene mutations and protein expression status, in understanding drug responses \cite{131}. In Table \ref{tab2}, we provide detailed case studies showcasing the applications of machine learning and deep learning regression models. Some models focus on specific tasks, such as single-task gene expression prediction, while others aim to generalize across multiple domains by training on diverse datasets, often leveraging self-supervised learning methods.
The primary objective of these models is to capture broadly applicable biological knowledge, enabling their use in various tasks, such as modeling nucleotide-peptide interactions or integrating cross-modal omics data \cite{132}. For instance, OmniBioTE \cite{52} utilizes transformer models to integrate nucleotide and peptide sequences, predicting biological phenomena such as binding free energy. These tasks are frequently accomplished through zero-shot inference or by adjusting model architectures—particularly the final layer—and fine-tuning for specific tasks.
To achieve these objectives, models enhance predictive performance by integrating multi-omics features with task-specific optimization strategies. Furthermore, some models facilitate task transfer through architectural adjustments or fine-tuning, enabling broader applicability and improved performance across diverse tasks.

Despite significant advancements in multi-omics regression using deep learning, several key challenges persist, including limited model interpretability, high computational complexity and constraints related to data scale \cite{129}. Among these, insufficient model interpretability hinders deeper understanding of biological mechanisms \cite{132}. To address this issue, Wang et al. incorporated attention mechanisms to quantify the contributions of gene mutations and RPPA data to drug response prediction, offering an effective approach to improving model interpretability \cite{131}.
To tackle the challenge of limited labeled data, Seal et al. proposed a semi-supervised learning framework that leverages information from unlabeled data, partially mitigating the impact of data scarcity on model performance \cite{129}. Furthermore, the high computational demands associated with integrating heterogeneous omics data remain a significant obstacle. To enhance computational efficiency, the MOMA method applies dimensionality reduction and data normalization, significantly improving model performance \cite{fr3}.
Looking ahead, advanced techniques such as resource optimization, semi-supervised learning, and lightweight architectures are expected to further address challenges related to data scarcity and computational complexity. These methods not only have the potential to improve model generalization but also significantly reduce computational costs, facilitating the broader application of multi-omics deep learning in precision medicine and drug discovery.

\begin{tiny}
\begin{longtable}{p{1.5cm}p{1cm}p{2.5cm}p{1.5cm}p{1.5cm}p{1.5cm}p{0.5cm}}

\caption{\textbf{Application of modern machine learning techniques in multi-omics regression tasks. We list them according to their categories.}}
\label{tab2} \\
\toprule
    \rowcolor[HTML]{C6CCD9}
         \textbf{Specific Method}&  \textbf{Model Categories}&  \textbf{Specific Regression tasks}&  \textbf{Evaluation Indicators}&  \textbf{Data Sources}&  \textbf{Venue}& \textbf{Year}\\ \midrule
         \rowcolor[HTML]{F2F3F6} 
         Seal et al. \cite{129} & MLP &  Regression prediction of gene expression in hepatocellular carcinoma & $R^2$: 0.97 & TCGA &  Genomics & 2022\\ \hline\rowcolor[HTML]{E1E4EB}
         Almutiri et al. \cite{132} & Deep Forest & Predicting drug response in cancer cell lines & $R^2$: 0.71 & CCLE & International Journal of Advanced Computer Science and Applications & 2023\\ \hline\rowcolor[HTML]{F2F3F6} 
         Multi-Omics Model and Analytics (MOMA) \cite{fr3} & RNN &  Performed regression predictions on the growth dynamics and molecular concentrations of Escherichia coli &  PCC: 0.54 & 
         Ecomics & Nature 
         communications & 2016\\ \hline\rowcolor[HTML]{E1E4EB}
         Wang et al. \cite{131} & DNN & Predict the response of cancer cell lines to anti-cancer drugs & $R^2$: 0.90 & CCLE & BMC 
         Bioinformatics & 2021\\ \hline\rowcolor[HTML]{F2F3F6} 
         OmniBioTE \cite{52} & Trans-former & Prediction of Gibbs free energy ($\Delta$G) changes in peptide-nucleotide binding interactions & PCC:0.82 & GenBank,
         Uniref100 &  arXiv & 2024\\ \bottomrule \rowcolor[HTML]{E1E4EB}

\end{longtable}
\end{tiny}
%%%

\subsection{Generative Tasks}\label{subsec33}
In multi-omics research, data processing faces several significant challenges. On one hand, the high dimensionality and complexity of the data make it difficult for traditional methods to uncover the underlying patterns of biological systems \cite{135}. On the other hand, data sparsity and limited sample sizes further constrain the learning and generalization capabilities of models \cite{138}.
Generative models offer solutions to these challenges with their unique ability to learn from noisy data distributions and generate new data samples that are statistically consistent \cite{53, 141, 142}. This capability makes generative models particularly valuable for addressing issues of sparse or limited multi-omics datasets \cite{72, 74}. Currently, the most prominent generative models in this field include Generative Adversarial Networks (GANs) \cite{93,101,102,107,94,57}, Variational Autoencoders (VAEs) \cite{97,99,100,103,104,108,58,59,48}, and diffusion models \cite{109,112,114,116,110,111,60}. 
For example, Ghebrehiwet et al. demonstrated the use of GANs in generating synthetic electronic health record (EHR) data, highlighting their role in producing synthetic patient data to improve diagnostic accuracy while protecting privacy in precision medicine \cite{136}. However, traditional GAN models often encounter challenges such as mode collapse and training instability. To address these issues, researchers have optimized model architectures, such as by incorporating the Wasserstein distance in Wasserstein GANs, which significantly enhances training stability \cite{96}.
Conversely, Sumathipala et al. applied network diffusion in predicting miRNA-disease associations, uncovering disease-disease relationships based on shared miRNAs \cite{112}. These advancements illustrate the diverse applications and potential of generative models in multi-omics research.

Another key function of generative models is data integration, which is essential for achieving a comprehensive understanding of biological systems in multi-omics research, particularly in the fusion of cross-modal data \cite{105,116}. By integrating diverse data types, generative models can reveal complex interactions across different layers of data, thereby enhancing downstream tasks such as disease prediction and biomarker discovery \cite{94,110,111,113,137}.
For example, Cheng et al. demonstrated the superior performance of Variational Autoencoders (VAEs) and Generative Adversarial Networks (GANs) in drug design and molecular generation tasks following data integration \cite{137}. Furthermore, these models not only integrate various omics data types but also improve data resolution. A notable example is soScope, which unveils subtle features of tissue structure, pushing the boundaries of multi-omics research \cite{133}.

Certain modalities of single-cell multi-omics data, such as transcriptomics and epigenomics, often exhibit high noise and low coverage. To uncover precise patterns in cell subtypes, states, and dynamic changes, higher-resolution spatial, temporal, or omics data and advanced analytical methods are essential. These requirements surpass the capabilities of traditional single-omics approaches \cite{139}.
Generative models have become indispensable in single-cell multi-omics research, finding applications in multimodal data integration, cross-modal generation, and dynamic process modeling \cite{95,101,100,103,108}. For instance, by integrating multimodal data and generating cross-modal samples, researchers can simulate cellular perturbation responses under various conditions, providing critical insights into dynamic changes in cell states \cite{108}.
Additionally, MichiGAN leverages decoupled designs and semantic manipulation of latent representations to generate high-quality single-cell data, shedding light on the cooperative mechanisms of gene regulatory networks \cite{101}. In dynamic process modeling, siVAE employs interpretable latent representations to elucidate the dynamic roles of key gene modules in cell differentiation trajectories and disease phenotypes \cite{103}.
Overall, the application of generative models in single-cell multi-omics is largely task-specific. By combining self-supervised learning (SSL) and conditional generation strategies, these models significantly enhance flexibility and applicability in single-cell multi-omics research \cite{139}.

Enhancing the interpretability of generative models has been an ongoing focus of research, as it enables a deeper understanding of biological processes and provides significant biological insights \cite{95,103}. For instance, siVAE improves model interpretability by incorporating feature-level embeddings, allowing researchers to identify and analyze genes and pathways that drive variations in cell types, disease states, or other biological conditions \cite{103}. Additional examples of data generation are provided in Table \ref{tab3}.

\begin{tiny}
\begin{longtable}{p{1.5cm}p{1cm}p{1.5cm}p{2cm}p{2.5cm}p{1cm}p{0.5cm}}
\caption{\textbf{Application of modern machine learning techniques in multi-omics generation tasks. We list them according to their categories.}} 
\label{tab3} \\
    \toprule
    \rowcolor[HTML]{C6CCD9}
         \textbf{Specific Methods} &  \textbf{Model Categories}&  \textbf{Data Sources}&  \textbf{Target Application}&  \textbf{Key Features}&  \textbf{Venue}& \textbf{Year}\\ \midrule \rowcolor[HTML]{F2F3F6}
         scAEGAN \cite{57} & GAN & SymSim,Mouse Hematopoietic Stem Cell Dataset,Human Pancreatic Islet Cell Dataset,scRNA-seq/scATAC-seq Paired Dataset & Single-Cell Data Integration &  Superior cross-modal integration & Plos one & 2023\\ \hline\rowcolor[HTML]{E1E4EB}
         Subtype-GAN \cite{93} & GAN & TCGA & Cancer subtyping & Integrative clustering & Bioinformatics & 2021\\ \hline\rowcolor[HTML]{F2F3F6}
         OmicsGAN \cite{94} & GAN & TCGA & Disease phenotype prediction & Integrates mRNA and miRNA with interaction networks & Bioinformatics & 2022\\ \hline\rowcolor[HTML]{E1E4EB}
         MichiGAN \cite{101} & GAN & Single-cell RNA-seq datasets & Cellular identity manipulation & Sampling from disentangled representations & Genome Biology & 2021\\ \hline\rowcolor[HTML]{F2F3F6}
         Nu{\ss}berger et al. \cite{102} & GAN/
         Auto-encoder & SNP data &  Data simulation under privacy constraints & Evaluates generative models for binary omics data with limited samples & Briefings in Bioinformatics & 2021\\ \hline \rowcolor[HTML]{E1E4EB}
         Al-Hurani et al. \cite{107} & GAN/
         Auto-encoder & TCGA & Cancer classification & class imbalance handling and classification & arXiv & 2024\\ \hline\rowcolor[HTML]{F2F3F6}
         Multi-Omics Integration Autoencoder \cite{58} & Auto-encoder & TCGA & Cancer Survival Prediction & Robust patient subgroup identification & Clinical Cancer Research & 2018\\ \hline \rowcolor[HTML]{E1E4EB}
         DeEPsnap \cite{59} & Auto-encoder & DEG (Database of Essential Genes), Ensembl,BioGRID,Gene Ontology,CORUM,Pfam & Essential Gene Prediction & High accuracy gene essentiality prediction & bioRxiv & 2024\\ \hline\rowcolor[HTML]{F2F3F6}
         SAEsurv-net \cite{48} & Auto-encoder & TCGA & Cancer Survival Prediction & Reduces data heterogeneity, improves accuracy &  arXiv & 2022\\ \hline \rowcolor[HTML]{E1E4EB}
         Boyeau et al. \cite{97} & Auto-encoder & PBMC & Differential gene expression analysis & estimation of log-fold changes in gene expression & bioRxiv & 2019\\ \hline\rowcolor[HTML]{F2F3F6}
         Hess et al. \cite{99} & Auto-encoder & cortical single-cell gene expression data & Pattern extraction & Uses log-linear models to extract patterns from synthetic data generated by deep generative models & Bioinformatics & 2020\\ \hline\rowcolor[HTML]{E1E4EB}
         DiffVAE \cite{100} & Auto-encoder & scRNA-seq & Modeling cell differentiation & Integrates graph autoencoders for enhanced analysis of cell relationships & Scientific Reports & 2020\\ \hline \rowcolor[HTML]{F2F3F6}
         siVAE \cite{103} & Auto-encoder & Fetal Liver Atlas & Genomic data analysis & Enhances interpretability of VAEs for genomic data, identifies gene modules and hubs & Genome Biology & 2023\\ \hline\rowcolor[HTML]{E1E4EB}
         Multi-Omics Generative Model \cite{104} & Auto-encoder & TCGA,SBIC & Cancer phenotype prediction & Uses padding to handle missing data, predicts cancer phenotypes, addresses class imbalance & AI & 2024\\ \hline \rowcolor[HTML]{F2F3F6}
         scCross \cite{108} & Auto-encoder & Matched Mouse Multi-Omics Datasets & Data integration, cross-modal generation & seamless integration of single-cell multi-omics data & Genome Biology & 2024\\ \hline\rowcolor[HTML]{E1E4EB}
         Soft Value-Based Decoding in Diffusion Models \cite{60} & diffusion & ZINC-250k, HepG2 cell line , 5'UTR stability and translational efficiency
dataset & Sequence Generation Optimization & Uses non-differentiable guidance & arXiv & 2024\\ \hline \rowcolor[HTML]{F2F3F6}
        Precious2GPT \cite{109} & Diffusion & LINCS L1000 & Synthetic multi-omics data generation & Combines CDiffusion and MoPT to generate multi-species, tissue-specific data & npj Aging & 2024\\ \hline\rowcolor[HTML]{E1E4EB}
         Network Diffusion \cite{110} & Diffusion & Diverse gene-centered omics datasets & Multi-omics integrative analysis & Amplifies gene associations within networks, used for precision medicine & Frontiers in Genetics & 2020\\ \hline \rowcolor[HTML]{F2F3F6}
         Multi-omics integration via weighted affinity and self-diffusion(MOSD) \cite{111} & Diffusion & TCGA & Cancer subtype identification & Assigns variable weights to omics, self-diffusion enhances patient clustering for subtype analysis & Journal of Translational Medicine & 2024\\ \hline\rowcolor[HTML]{E1E4EB}
         Network Diffusion (MAP Method) \cite{112} & Diffusion & miRTarBase DisGeNET & miRNA-disease relationship prediction & Uses network diffusion on interactome data to uncover miRNA roles in complex diseases & Scientific Reports & 2020\\ \hline \rowcolor[HTML]{F2F3F6}
         His-MMDM \cite{114} & Diffusion & TCGA & Histopathology image translation & Translates images across tumor types and omics profiles & medRxiv & 2024\\ \hline\rowcolor[HTML]{E1E4EB}
         Network-based Integration of Multi-omics Data(NetICS) \cite{116} & Diffusion & TCGA &  Cancer gene prioritization & Uses bidirectional diffusion for ranking genes in cancer, integrating genetic and expression data &  Bio-informatics & 2018\\ \bottomrule \rowcolor[HTML]{F2F3F6}

\end{longtable}
\end{tiny}
%%%

\subsection{Clustering Tasks}\label{subsec34}
Early multi-omics data analysis predominantly relied on statistical and traditional machine learning methods, such as Principal Component Analysis (PCA) \cite{61,62}, Linear Discriminant Analysis (LDA), and Non-negative Matrix Factorization (NMF) \cite{63}. Traditional machine learning models have also demonstrated notable success in addressing relatively simple high-dimensional multi-omics data. For example, Zhang et al. utilized an enhanced PCA method (PCA-Plus), adapted from PCA, to analyze multi-omics data from TCGA, achieving p-values of less than 0.0005 across three datasets, which highlighted its effectiveness in dimensionality reduction \cite{62}.
While modifications to traditional machine learning models enable them to handle some nonlinear relationships, these approaches still struggle to capture complex nonlinear interactions and effectively manage inter-group heterogeneity in multi-omics data. Consequently, they are limited in both interpretability and adaptability, which restricts their broader application.

\begin{tiny}
\begin{longtable}{p{2cm}p{1.5cm}p{2.5cm}p{2.5cm}p{1cm}p{0.5cm}}
\caption{\textbf{Application of modern machine learning techniques in multi-omics clustering tasks. We list them according to their categories.}}
\label{tab4} \\
\toprule \rowcolor[HTML]{C6CCD9}
        \textbf{Specific Methods}&  \textbf{Model Categories}&  \textbf{Data Sources}&  \textbf{Specific Application}&  \textbf{Venue}& \textbf{Year}\\ \midrule \rowcolor[HTML]{F2F3F6}
        Poisson Correction Distance-based Embedding
        (PCD2Vec) \cite{61} & PCA & COVID-19 spike protein sequences from various hosts & Dimensionality reduction for host classification of COVID-19 spike proteins & IJCNN & 2023\\ \hline\rowcolor[HTML]{E1E4EB}
        PCA-Plus \cite{62} & PCA &  TCGA & Dimensionality reduction for batch effect analysis & bioRxiv & 2024\\ \hline\rowcolor[HTML]{F2F3F6}
        Stacked Autoencoder \cite{48} & Auto-encoder & TCGA & reduce liver cancer multi-omics data to key survival features & arXiv & 2022\\ \hline\rowcolor[HTML]{E1E4EB}
        Deep Learning Autoencoder \cite{58} & Auto-encoder & TCGA &  compress multi-omics data into 100 features in the bottleneck layer &  Clinical Cancer Research & 2018\\ \hline\rowcolor[HTML]{F2F3F6}
        CLCluster \cite{116} & Auto-encoder &  TCGA & Cancer subtype identification and alternative splicing analysis & bioRxiv & 2024\\ \hline\rowcolor[HTML]{E1E4EB}
        Single Cell Multimodal Deep Clustering(scMDC) \cite{118} & Auto-encoder/K-means & CITE-seq and SMAGE-seq datasets & Clustering multimodal single-cell data, improving cell type identification and addressing batch effects & Nature Communications & 2022\\ \hline\rowcolor[HTML]{F2F3F6}
        SpaHDmap \cite{63} & Non-negative matrix factorization (NMF)/ GNN & Mouse Brain Spatial Transcriptomics Dataset, PDOX (Patient-Derived Orthotopic Xenograft) Mouse Model Dataset, Human Colorectal Cancer Dataset & Dimensionality reduction of spatial transcriptomics and histology images for high-resolution embedding & bioRxiv & 2024\\ \hline\rowcolor[HTML]{E1E4EB}
        Graph Regularized Multi-view Ensemble Clustering for Single-Cell(GRMEC-SC) \cite{117} & Non-negative matrix factorization(NMF) &  Single-cell multi-omics datasets & Clustering single-cell data and identifying cell heterogeneity & Bioinformatics & 2024\\ \hline\rowcolor[HTML]{F2F3F6}
        Transformer with Convolution and Graph-Node co-embedding (TCGA) \cite{64} & GNN & HER2-positive breast tumor spatial transcriptomics dataset & reduce dimensionality by extracting key features from histopathological images for gene expression prediction & Medical Image Analysis & 2024\\ \bottomrule

\end{longtable}
\end{tiny}

Deep learning has significantly advanced dimensionality reduction for complex, nonlinear data. Generative models, such as autoencoders (AEs) and variational autoencoders (VAEs), have demonstrated strong performance in this domain. For example, Kumardeep Chaudhary et al. applied an autoencoder to multi-omics hepatocellular carcinoma (HCC) data from TCGA, achieving a C-index of 0.68, an improvement of 0.06 over PCA \cite{58}. Similarly, graph neural networks (GNNs) have shown effectiveness for graph-structured data \cite{64}. However, both generative models and GNNs face challenges, including information loss and unstable training, which limit their ability to fully capture complex relationships and reduce the quality of dimensionality reduction.

To address these issues, combining convolutional neural networks (CNNs) and transformers has proven promising. CNNs are adept at extracting local spatial features, while transformers leverage self-attention mechanisms to capture global dependencies. This combination improves feature extraction and stability. For instance, Xiao et al. proposed the TCGN (Transformer with Convolution and Graph-Node co-embedding) model, which integrates CNNs and transformers. TCGN demonstrated superior performance in gene expression prediction, achieving a Pearson correlation coefficient (PCC) of 0.1303—nearly a tenfold improvement over the ST-Net model \cite{64}.
Despite these advances, CNNs and transformers still tend to focus on either local or global feature extraction and face challenges in capturing complex relationships, as well as issues with training efficiency and model scalability. Introducing larger models that better integrate local and global information, coupled with pretraining strategies, could enhance generalization and overall effectiveness.

In multi-omics clustering tasks, contrastive learning-based models, such as CLCluster, have proven effective, particularly in cancer subtype recognition. By combining contrastive learning with mean-shift clustering, these models achieve high adaptability and accuracy \cite{117}. For single-cell multi-omics data, approaches like GRMEC-SC integrate GNNs with multi-view ensemble learning, enhancing clustering accuracy and stability. This method excels by combining diverse omics data at the single-cell level, demonstrating competitive performance across complex datasets \cite{118,119}.
These innovations mark significant advancements in clustering and dimensionality reduction for multi-omics data by improving model adaptability, accuracy, and biological relevance.
 
\section{Research Progress in Multi-Omics Fundamental Models}\label{sec4}

The foundation model has recently become a hot topic, referring to the self-supervised learning mechanism of large models designed to perform masked learning on large-scale biological sequences. These models aim to learn the ``language of biology''. Specifically, the process often involves randomly masking 15\% of the sequence's information and using a language model for in-context learning to capture the contextual relationships within the sequence. The model's effectiveness is then validated through various computational biology tasks, establishing it as a universal model. Common foundation models are based on architectures such as ELMo \cite{65}, Transformer \cite{66}, BERT \cite{67}, and BART \cite{68}. However, since the Bi-LSTM module in ELMo lacks the ability to process long-range dependencies in sequences, models are often built using the latter three architectures to analyze biological sequence information spanning billions of characters.

\begin{figure}[tbp]
    \centering
    \includegraphics[width=1\textwidth]{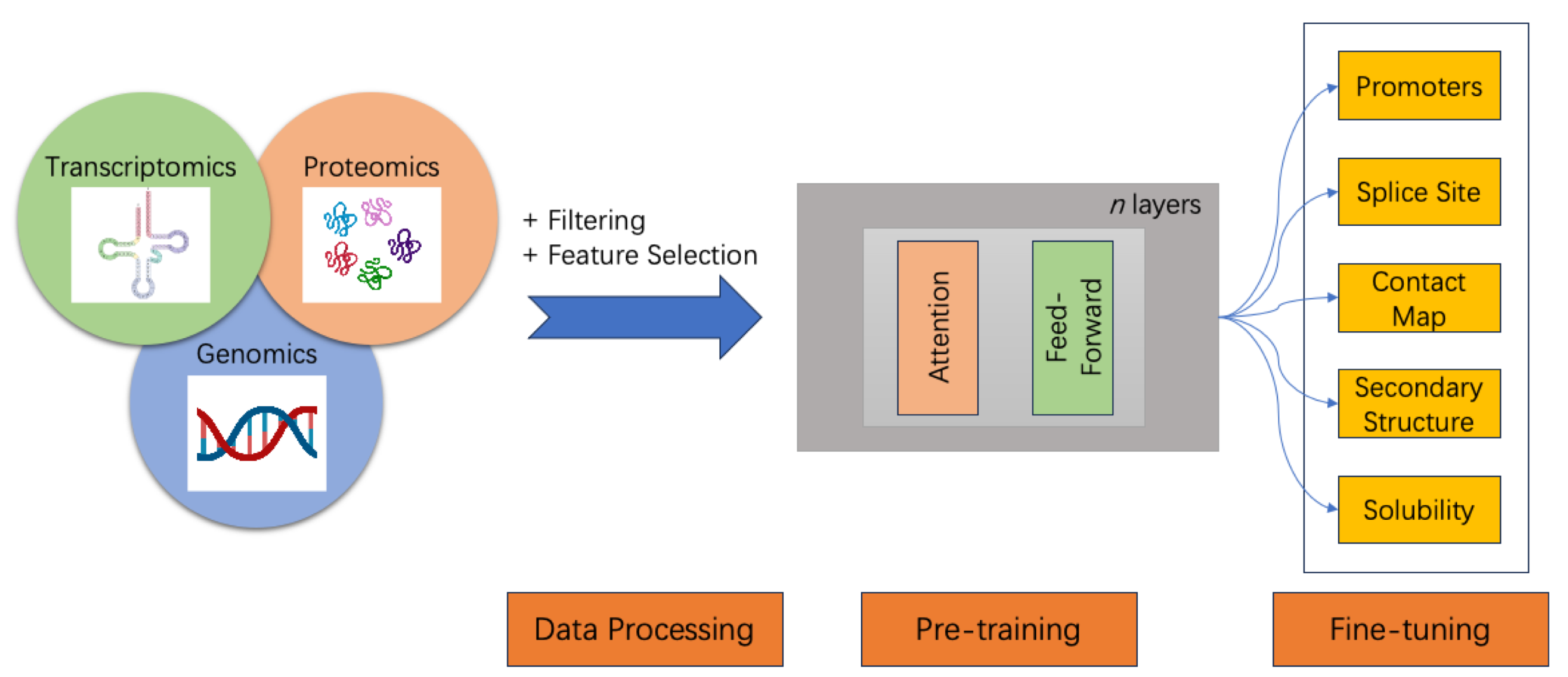}
    \caption{\textbf{Flowchart of the Multi-Omics Foundation Model.} 
    The training paradigm of the multi-omics foundation model consists of two phases. The first phase, the pre-training phase, focuses on learning the semantic information embedded in biological sequences. The second phase, the fine-tuning phase, is designed to adapt the model to specific downstream omics tasks.}
    \label{fig2}
\end{figure}

When designing multi-omics models, it is essential to incorporate biological prior knowledge or validate whether the model can understand and utilize this knowledge. To develop interpretable foundational models for multi-omics, it is necessary to integrate biological prior knowledge while creating an efficient mechanism for cross-modal information fusion, enabling seamless interaction between different biological macromolecules.
Evo \cite{69} represents the first attempt at a foundational model for multi-omics tasks. It combines the Hyena layer with the Rotational Position Encoding (RoPE) multi-head attention mechanism to support long-sequence modeling. Similarly, scGPT \cite{70}, pre-trained on a single-cell dataset comprising over 33 million samples, has achieved state-of-the-art performance in various tasks, including cell annotation, multi-batch integration, multi-omics integration, perturbation response prediction, and gene network inference. However, these methods primarily demonstrate the foundational model’s ability to handle multi-omics tasks through downstream applications.

To better evaluate a foundational model’s capacity to process DNA, RNA, and protein interactions, LucaOne \cite{71} introduces a DNA-protein classification task, providing a more direct assessment of the model’s performance in multi-modal classification scenarios. Furthermore, recognizing the importance of incorporating biological prior knowledge on multi-molecular interactions during the pre-training phase, CD-GPT \cite{72} establishes a multi-omics large model using a two-stage pre-training approach. In the first stage, it pre-trains on individual molecules, followed by a second stage where it pre-trains on multiple molecules while allocating weights effectively.

\section{Conclusion and Outlook}\label{sec5}
We have reviewed the current applications of artificial intelligence methods in multi-omics tasks. To conclude, we will highlight a few emerging topics in this field.

\subsection{Development of A Cross-Species Evaluation System}\label{subsec51}
Multi-omics integration is a prominent research topic in computational biology, with widespread applications across various biological tasks, diseases, and species. However, due to the selective gene expression observed in different biological species and cell types, foundational models pre-trained in specific areas often fail to generalize effectively to other tasks. This underscores the critical need for task-oriented multi-omics evaluation benchmarks.
For instance, Hu et al. \cite{73} developed an evaluation benchmark for single-cell multi-omics prediction and integration, highlighting the superior performance of Seurat, MOJITOO, and scAI in downstream single-cell multi-omics tasks. Similarly, Mangnier et al. \cite{74} introduced an evaluation benchmark based on gut microbiome metagenomics and metabolomics data. In the field of cancer research, Yang et al. \cite{75} proposed Subtype-Former, a cutting-edge tool for cancer dimensionality reduction that significantly outperformed other benchmark models. Additionally, CBOW \cite{76} offers a user-friendly cancer multi-omics evaluation benchmark, incorporating over 20 cancer-related tasks and providing organized model scripts.

Furthermore, evaluating methods for ensuring the privacy and security of multi-omics data is equally critical. Pfeifer et al. \cite{77} designed a multi-omics privacy protection model using joint unsupervised random forests and benchmarked it against other models to demonstrate its effectiveness.

\subsection{Development of Interpretable Models Centered on Biological Rules}\label{subsec52}
Efficient integration of multi-omics data remains the most significant challenge for foundational models in this field. Current integration methods primarily rely on matrix concatenation \cite{71} or cross-attention mechanisms \cite{78}. To achieve better alignment of information across different modalities, advanced multimodal fusion strategies should be incorporated into multi-omics tasks.
For instance, the Q-Former structure in BLIP-2 \cite{79}, a lightweight Transformer architecture, has proven effective for integrating information across modalities and has already been applied to RNA molecule generation tasks. In the future, emerging architectures such as Cambrian-1 \cite{80}, MM-LLMs \cite{81}, Honeybee \cite{82}, and DECO \cite{83} should also be explored for their potential in addressing multi-omics challenges.

Moreover, integrating biological knowledge into multi-omics tasks is crucial. CD-GPT \cite{72} exemplifies this approach by considering interactions among different biological macromolecules involved in the central dogma. It employs a two-stage modeling process, enhancing the model’s understanding of multi-modal biological sequences during the second stage.

\subsection{Long Sequence Modeling}\label{subsec53}

Traditional biostatistical models often focused on analyzing specific discrete sites within biological sequences, overlooking the long-range interactions between genes. Current biological sequence models are predominantly based on spatial state equations \cite{84,85} or Transformer architectures \cite{86,87}. However, the length of biological sequences often reaches billions of characters, posing significant challenges.
For instance, TTT \cite{88}, a long-text model with linear complexity, was the first to extend processing sequence lengths to over 16k, significantly enhancing the understanding of biological sequences. Nevertheless, modeling large-scale biological sequences requires substantial memory and computational resources. To address this, reducing model complexity and adopting efficient parallel computing strategies are critical considerations in model design.
For example, InstInfer \cite{89} accelerates inference by offloading the attention mechanism and kv cache to computer storage drives. Similarly, Mnemosyne \cite{90} introduces a 3D parallel scheme that achieves efficient inference with up to 10 million contexts in approximately 30 milliseconds.

\subsection{Release of The Multi-Organ Dataset}\label{subsec54}

When integrating data from different sources or platforms, batch effects are a common challenge. To minimize the impact of batch effects during the model pre-training process, data from various organs has been collected for numerous tasks. For instance, the dataset used in the pre-training phase of scFoundation \cite{91} comprises over 50 million single-cell samples from different tissues and organs. However, the availability of multi-omics data remains limited. Due to the selective expression of genes, publicly available datasets often focus on single-organ data related to specific diseases \cite{75}. To advance the development of large-scale multi-omics models, it is crucial to make more multi-omics databases from diverse organs publicly accessible in the future.

\subsection{Frontier Issues in Biological Migration}\label{subsec55}
In single-cell research, there is a growing trend of utilizing multi-tissue single-cell sequencing data to construct cell atlases, enabling the identification of conserved cell functions and uncovering how cells adapt to their environments. For instance, Rosen et al. \cite{92} developed a universal cell embedding based on 36 million single-cell data points from eight species, effectively identifying previously uncharacterized species in the cell atlas. Additionally, CELLama \cite{39} transforms cell data into "sentences" that encode gene expressions and metadata, facilitating cell type identification and revealing complex cell interactions within the cell atlas. SCimilarity \cite{40} employs metric learning to create interpretable unified representations of cell types, allowing rapid queries for cells with similar morphologies.

Single-cell sequencing methods, such as single-cell RNA sequencing (scRNA-seq), have been extensively applied to predict immunotherapeutic responses \cite{c5add1}. For example, comparative analysis of two single-cell transcriptome datasets from basal cell carcinoma (BCC) and pancreatic ductal adenocarcinoma (PDAC) has revealed differences in immune mechanisms involved in tumor progression between these two cancers \cite{c5add2}. Single-cell sequencing can also identify known antigen-specific sequences, though this technology remains in its infancy. Advances in deep sequencing have significantly expanded the available data on the adaptive immune receptor repertoire (AIRR). In response, the AIRR Community has been established to standardize the outputs of immune repertoire sequencing experiments and manage immune repertoire sequencing data systematically \cite{c5add3,c5add4}.

Despite these advancements, the application of ML in immunology remains relatively immature and underdeveloped. Existing methods are often simplistic (Table \ref{tab1}), unable to comprehensively analyze data, and hindered by a severe lack of immune repertoire datasets \cite{c5add5}. To address these challenges, it is essential to develop advanced experimental techniques, improve the integration of experimental datasets through ML and deep learning (DL) methods, and refine immune and antigen functional databases. These efforts would enhance the determination of immune status, prediction of antigen specificity, and the design and screening of immune receptor candidate drugs.

Machine learning and deep learning models offer the potential to identify complex, high-dimensional patterns in immune repertoire data. Biomarkers derived from ML methods can be employed to predict clinical outcomes of immunotherapy and support personalized immunotherapy approaches \cite{c5add6}. Therefore, refining ML and DL methods is critical to advancing immunotherapy and promoting its broader application in cancer treatment and precision medicine.

\bibliography{sn-bibliography.bib}% common bib file
% \input sn-article.bbl
\end{sloppypar}
\end{document}